\documentstyle[11pt,aaspp4,flushrt]{article}
\def\eg{{\it e.g.~}}

\def\Sec{\rlap{$^{\prime\prime}$}.\hbox to 2pt{}}
\def\Min{\rlap{$^\prime$}.\hbox to 1pt{}}

\input epsf
\begin{document}

\title{HST Imaging of the Globular Clusters
in the Fornax Cluster: Color and Luminosity
Distributions\footnote[1]{Based on observations with the NASA/ESA
Hubble Space Telescope, obtained at the Space Telescope Science
Institute, which is operated by AURA, Inc., under NASA contract NAS
5-26555.}}

\author{Carl J. Grillmair}
\affil{SIRTF Science Center, California Institute of Technology,
Mail Stop 100-22, Pasadena, California 91125\\
{\it email}: carl@ipac.caltech.edu}

\author{Duncan A. Forbes} \affil{School of Physics and Astronomy,
University of Birmingham, Edgbaston\\ Birmingham B15 2TT, United
Kingdom\\ {\it email}: forbes@star.sr.bham,ac.uk}

\author{Jean Brodie} \affil{Lick Observatory, University of
California, Santa Cruz, California, 95064\\
{\it email}: brodie@ucolick.org}

\author{Rebecca Elson} \affil{Institute of Astronomy, Madingley Road,
Cambridge CB3 0HA, England\\
{\it email}: elson@ast.cam.ac.uk}

\begin{abstract}

We examine the luminosity and $B-I$ color distribution of globular
clusters for three early-type galaxies in the Fornax cluster using
imaging data from the Wide Field/Planetary Camera 2 on the {\it Hubble
Space Telescope}.  The luminosity functions we derive are in most
cases better than 50\% complete down to $B = 26.6$.  We find that the
color distributions of globular clusters in the central region of NGC
1399 and its nearby neighbor NGC 1404 are bimodal and statistically
indistinguishable. The metallicity spread, as inferred from the color
distributions in these two galaxies, is very similar to that of M 87.
NGC 1399's luminosity function is also very similar to that of M 87,
and comparing their respective peak magnitudes indicates that the
Fornax cluster is at very nearly the same distance from the Local
Group as is the Virgo cluster. From this we derive $H_0 = 82 \pm 8$ km
s$^{-1}$ Mpc$^{-1}$, where the uncertainty reflects only the effects
of random errors. The number of unresolved objects we find at a
projected distance of 440 kpc from NGC 1399 is consistent with nothing
more than compact background galaxies, though the small field of view
of the WFPC2 does not allow us to put strong constraints on the number
of intergalactic globular clusters.  The luminosity function of
objects detected around NGC 1316 is more nearly exponential than
log-normal, and both the color and size distribution of these objects
distinguishes them from the clusters surrounding NGC 1399. We suggest
that these objects are more akin to old open clusters in the Galaxy
than they are to globular clusters in typical early-type galaxies.

\end{abstract}


\section{Introduction.}

The characterization of the globular cluster luminosity function
(GCLF) remains a high priority as a means to better understand both
the formation of globular clusters and the buildup of galaxies. The
differences in the breadth and peak position of GCLFs in late-type
galaxies in the Local Group and early-type galaxies in other clusters
has been known for some time (Harris et al. 1991). Efforts to use GCLFs
as distance indicators or tracers of galaxy mergers would clearly
benefit from more complete coverage over luminosity and galaxy
type. With its 0\Sec1 resolution, low sky background, and consequent
ability to distinguish very faint point-sources, the {\it Hubble Space
Telescope} ({\it HST}) is an ideal platform from which to carry out
such a study.

The first surveys of globular clusters in the Fornax cluster were
carried out photographically (Dawe \& Dickens 1976; Hanes \& Harris
1986; Harris \& Hanes 1987).  Later CCD studies (see Ashman \& Zepf
1997 for a list of references) detected bimodal color distributions
(Zepf \& Ashman 1993) and color gradients (Ostrov, Geisler, \& Forte
1993) in the cluster system of NGC 1399. {\it HST} studies have now
confirmed the existence of multi-modal color distributions in several
other giant ellipticals (e.g. M 87, Elson \& Santiago 1996b; Whitmore
et al. 1995; NGC 5846, Forbes et al. 1997a).  What follows is the first
systematic {\it HST} investigation of the globular cluster systems in
the Fornax cluster.  The results should help us to understand how cD
galaxies and their large associated globular cluster populations
formed (Ashman \& Zepf 1992; Forbes, Brodie, \& Grillmair 1997b).

One aim of this study is to search for variations in the GC
populations among a sample of early-type galaxies in the same
cluster. In this paper we examine the globular cluster populations of
NGC 1399, NGC 1404, and NGC 1316. In companion papers we study the
metallicity and spatial distributions of globular clusters in NGC 1399
and NGC 1404 (Forbes et al. 1998) and examine the globular cluster
system of the lower luminosity elliptical NGC 1379 (Elson et al. 
1998). NGC 1399 is the cD elliptical at the center of the Fornax
cluster and, like M 87, has an extraordinarily large
 number of globular clusters for its luminosity (see references in
Ashman \& Zepf 1997). NGC 1404 is an E1 elliptical about half a
magnitude fainter and lying only 10 arcminutes from NGC 1399 in
projection. Attempts to study its globular cluster system (Hanes \&
Harris 1986; Richtler et al. 1992) have been complicated by its
location within the extended envelope of NGC 1399. NGC 1316 is another
giant elliptical in the Fornax cluster, more than a magnitude brighter
than NGC 1399, with a double-lobed radio source (Fornax A), an X-ray
halo, and dust and other irregularities (Schweizer 1980) indicating a
recent merger. NGC 1316 is situated more than three degrees away from
NGC 1399, clearly dominating its corner of the Fornax cluster, and has
been the subject of a recent WFPC1 study by Shaya et al. (1996).

If the high specific frequency of globular clusters around NGC 1399 is
to be attributed to tidal stripping or accretion of existing globular
clusters in other cluster galaxies (which would be consistent with the
kinematic data of Grillmair et al. 1994a and Kissler-Patig et al. 
1997b), we might expect mean GC metallicities to be very similar to
those of other cluster galaxies. Alternatively, if much of NGC 1399's
globular cluster system formed more recently from enriched gas, its
globulars may be overly metal-rich compared to other galaxies of the
same magnitude. If new globular clusters are formed in the normal
course of mergers (Ashman \& Zepf 1992), then we might expect to see a
a number of young, metal-rich globulars surrounding NGC 1316.

We describe the observations and their analysis in Section
\ref{sec:obs}. The color distribution of detected globular
clusters is discussed in Section \ref{sec:color}. We analyze and
compare the luminosity functions in Section \ref{sec:lf}. Section
\ref{sec:n1316} more closely examines the characteristics of objects
around NGC 1316, and Section \ref{sec:inter} briefly discusses the
possibility of intergalactic globular clusters in our sample. We
summarize our conclusions in Section \ref{sec:summary}. The
metallicity and spatial distributions of globular clusters in these
galaxies and their implications for formation scenarios are discussed
by Forbes et al. (1998).

\section{Observations and Photometry. \label{sec:obs} }

An observing log for program GO \#5990 is given in Table 1.  While NGC
1379 was observed as an example of a ``normal'' elliptical, telescope
acquisition problems prevented useful F814W data from being taken by
the time of writing.  The NGC 1379 observations in the F450W filter
are discussed by Elson et al. (1998).  Field 0338 (F0338) is situated
in the cD envelope of NGC 1399, at approximately the same projected
distance from NGC 1399 as NGC 1404, but on the opposite side of the
galaxy.  Field 0336 (F0336) is situated approximately 1.4 degrees
south of NGC 1399 in a blank region of sky and serves to measure the
surface density of background sources. The relative positions and
orientations of these fields are shown in Figure \ref{fig:plate1}.
Filters F450W ($\sim B$) and F814W ($\sim I$) were chosen to increase
metal sensitivity by about a factor of two over that attainable using
$B - V$ (Brodie 1981; Geisler, Lee, \& Kim 1996). The images were
ADC-corrected, bias-subtracted, and flat-fielded in the course of
standard pipeline processing.

\begin{figure}[tb]
\caption{Digitized Sky Survey image of the Fornax cluster showing
positions and orientations of three of the four WFPC2 pointings
discussed in this paper. The NGC 1379 pointing is discussed separately
by Elson et al. (1997). The entire field shown here subtends $2^\circ$
on a side.  \label{fig:plate1}}
\end{figure}

\begin{deluxetable}{lcccccc}
\tablecaption{Observing Log.}
\tablecolumns{7}
\tablehead{
\colhead{Target} &
\colhead{RA} &
\colhead{Dec} &
\colhead{Date} &
\colhead{Filter} &
\colhead{Exposure Times} &
\colhead{Orientation\tablenotemark{a}} \\
\colhead{} &
\multicolumn{2}{c}{J2000} &
\colhead{} &
\colhead{} &
\colhead{s} &
\colhead{$^o$} \\
}
\startdata
NGC 1399 & 03 38 29.0 & -35 27 00.5 & 1996 Jun 2 & F814W & 3 $\times$ 600 & 242.0 \nl
NGC 1399 &  &  & 1996 Jun 2 & F450W & 4 $\times$ 1300 & 242.0 \nl
NGC 1379 & 03 36 03.9 & -35 26 26.0 & 1996 Mar 11 & F814W & 3 $\times$ 0.5 & 148.3 \nl
NGC 1379 &  &  & 1996 Mar 11 & F450W & 2 $\times$ 1300 + 2 $\times$ 500 + 1400 & 148.3 \nl
NGC 1404 & 03 38 51.7 & -35 35 36.0 & 1996 Apr 3 & F814W & 260 + 600 + 1000 & 171.8 \nl
NGC 1404 &  &  & 1996 Apr 3 & F450W &  2 $\times$ 500 + 2 $\times$ 1300 + 1400 & 171.8 \nl
NGC 1316 & 03 22 41.8 & -37 12 29.8 & 1996 Apr 7 & F814W & 260 + 600 + 1000 & 181.3 \nl
NGC 1316 &  &  & 1996 Apr 7 & F450W &  2 $\times$ 500 + 2 $\times$ 1300 + 1400 & 181.3 \nl
Field 0338 & 03 37 57.0 & -35 21 54.5 & 1996 Apr 6 & F814W & 3 $\times$ 600 & 175.3 \nl
Field 0338 & & & 1996 Apr 6 & F450W & 4 $\times$ 1300 & 175.3 \nl
Field 0336 & 03 36 02.7 & -36 45 54.0 & 1996 Apr 11 & F814W & 3 $\times$ 600 & 182.4 \nl
Field 0336 & & & 1996 Apr 11 & F450W & 4 $\times$ 1300 & 182.4 \nl
\enddata

\tablecomments{}
\tablenotetext{a}{Angle, measured North through East, of the y-axis of the PC chip.}
\end{deluxetable}

Each set of between 3 and 5 exposures of a given field was split into
two pointings offset by 0\Sec5 from one another. This corresponds
approximately to integer pixel shifts in both the PC (0\Sec0455
pix$^{-1}$) and the WF chips (0\Sec0996 pix$^{-1}$) (the actual shifts
are generally within 0.1 pixels of integer values). Aligning and
median-combining the images consequently allowed us to eliminate the
majority of cosmic rays and hot pixels at the same time. To prevent as
much as possible the contamination of the faint end of the GCLF, known
warm/hot pixels, charge traps, and bad columns were flagged and
subsequently ignored in the median procedure.  The distribution of
unflagged pixels was further tested for the presence of cosmic ray
events before computing the median value from the remaining pixels.
The results were found to be cleaner both statistically and visually
when compared with the results of such standard cosmic-ray removal
procedures as are available, for example, within STSDAS/IRAF.

The PC images of the sample galaxies were analyzed using the VISTA
surface-brightness profile-fitting routine SNUC, which fits
logarithmically-spaced, concentric ellipses to unmasked portions of
the galaxy profile. The resulting model profiles were then subtracted
from each image in an effort to flatten the background as much as
possible. For the WF chips, similar model surface brightness
distributions were generated using both large-scale window-medianing,
and convolution with a two-dimensional Gaussian having $\sigma =
1\Sec0$. Luminosity fluctuations in F0336 and F0338 were found to be
on scales sufficiently large that model subtraction was deemed
unnecessary.

Source detection and aperture photometry were carried out using
DAOPHOT II (Stetson 1987). Owing to the high resolution inherent in
these images, overlap of sources was negligible and aperture photometry
yielded a slightly narrower color distribution than could be obtained using
PSF-fitting photometry.  The detection threshold was fixed at three
times the rms expected {\it locally} from photon statistics and
readout noise, the former computed by taking into account the
subtracted model surface-brightness distributions. Given the degree of
undersampling inherent to the WFPC2 detectors, a $3\sigma$ detection
threshold predictably yielded a substantial number of spurious
detections. However, the final photometry table was generated using
only those sources which were detected in both passbands and having
colors in the range $-1.0 < B-I < 4.0$ . Consequently, the number of
spurious detections remaining in the final sample (as determined by
visual inspection) was found to be negligible.

NGC 1316 is something of a special case in our sample owing to the
presence of large amounts of dust obscuration. For this galaxy, we
masked large areas of the WFPC2 field of view in which there was
visible evidence of dust in F450W. The extent of the dust and the
regions we chose to exclude from analysis are shown in Figure
\ref{fig:plate2}. The ``unobscured'' detections attributed to NGC 1316
thus come from approximately the outer halves of each of the WF chips;
detections in the PC are not considered in the subsequent analysis
owing to the strong likelihood that all detections are obscured and
reddened to a significant extent.  

\begin{figure}[tb]
\caption{The WFPC2 F450W image of NGC 1316 divided by a model
surface-brightness distribution. The demarcation is that beyond which
clusters were deemed {\it not} to be seriously affected by extinction
and reddening. Note the interesting, radially-oriented columns of dust
in WF4, which suggest Rayleigh-Taylor instabilities or erosion by
particles or radiation from the nucleus of the galaxy.
\label{fig:plate2}}
\end{figure}

Faint background galaxies and bright foreground stars often produced
multiple detections along spiral arms or diffraction spikes, and
several small regions of each image were masked accordingly. In
addition, we disregarded all detections with $m_1 - m_2$ (the
difference in magnitude computed for apertures of radius 1 and 2
pixels, respectively) $ < -0.9$ magnitudes ($< -1.1$ magnitudes for
the PC). Visual examination showed that this effectively removed
almost all obvious faint galaxies from the sample.  A more extensive
discussion of the contamination issue is presented by Elson et al.
(1998), and as we show presently, the number of background objects in
our sample to $B = 26.5$ is $< 2\%$.

At the distance of the Fornax cluster, typical globular clusters are
marginally resolved by WFPC2. Treating them as point sources from the
standpoint of photometry would therefore underestimate to varying
degrees their total luminosities (see Holtzman et al. 1996 for a
discussion of this problem). To estimate this shortfall, we convolved
an array of King models (King 1966) with WFPC2 PSFs generated using
Tiny Tim (Krist 1995) for each globular's position on the detector. We
then sampled these models using WFPC2-sized pixels and a $5 \times 5$
grid of pixel-centerings. Each model realization was then compared on
a pixel-by-pixel basis with the real data in the NGC 1399 images for
$r \le 3$ pixels to compute $\chi^2$.

\begin{figure}[ht]
\epsfxsize=4.0in
\epsfbox[-96 72 504 700]{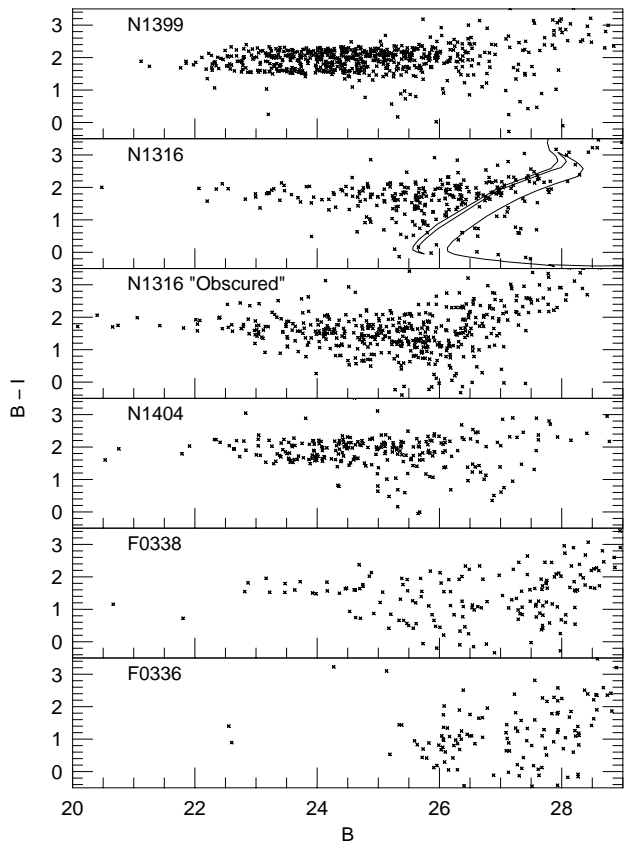}
\caption{Colors and magnitudes for all compact sources in the WFPC2
frames. An electronic tabulation of the photometry is available on
request from CJG. The solid line in the NGC 1316 panel shows the
color-magnitude sequence (subgiant and horizontal branch) expected for
a stellar population of age $4 \times 10^7$ yrs and [Fe/H] =
-0.4. \label{fig:cm}}
\end{figure}

While this procedure yielded a range of best-fit core radii not unlike
that found in Galactic globular clusters, it failed sufficiently often
that we do not regard the computed core radii as reliable. The reasons
for this may include the fact that we did not take into account image
dithering and subsequent integer-pixel shifting used to combat hot
pixels, as well as inaccuracies in the Tiny Tim model PSFs.  However,
aperture corrections derived from King model fits with acceptable
values of $\chi^2$ agreed very well with those determined empirically
for the brightest, most isolated clusters in the field. We therefore
adopted mean 0\Sec5 aperture corrections to our 2-pixel-radius
measurements determined directly from the images in each frame,
ranging from $\sim 0.25$ magnitudes for the WF chips to $\sim 0.65$
magnitudes for the PC. Based on our simulations, the use of mean
aperture corrections applied to all clusters in a given WF chip
is likely to introduce a random error of $\approx 0.1$ magnitudes to the
photometry of any individual cluster. To convert to $B$ and $I$
magnitudes we used the gain ratios and zeropoints given by Holtzman et
al. (1995; 1997).  The final magnitudes and colors of all detected
objects are plotted in Figure \ref{fig:cm}, and are available in
electronic form from CJG.

\begin{figure}[ht]
\epsfxsize=4.0in
\epsfbox[-72 36 528 536]{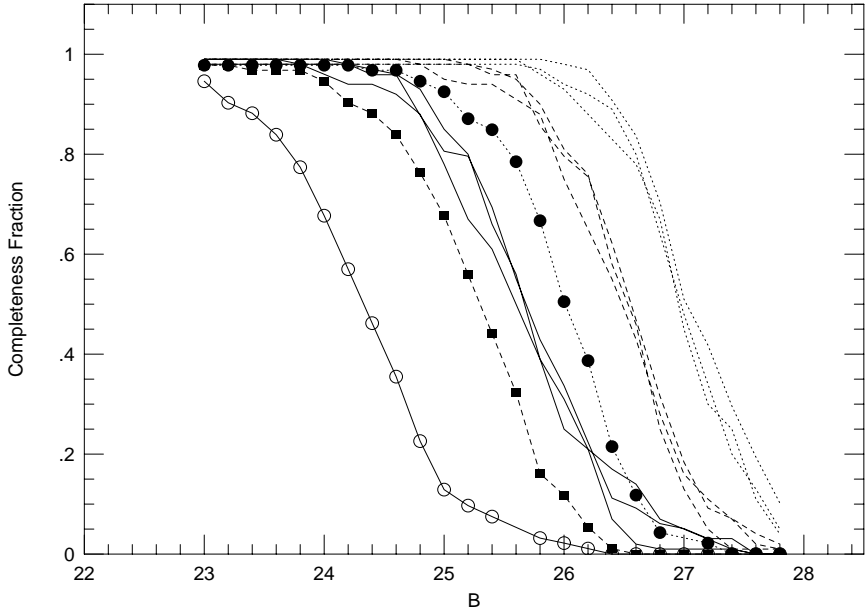}
\caption{Completeness fraction as a function of {\it B} magnitude
for NGC 1399.  Solid lines show the results for $B - I = 0.3$, dashed
lines correspond to $B - I = 1.3$, and dotted lines to $B - I = 2.3$.
Lines with symbols show the results for the PC, while all others
indicate completeness fractions for the WF chips. \label{fig:compfig}}
\end{figure}

\begin{figure}[ht]
\epsfxsize=4.0in
\epsfbox[-72 36 528 536]{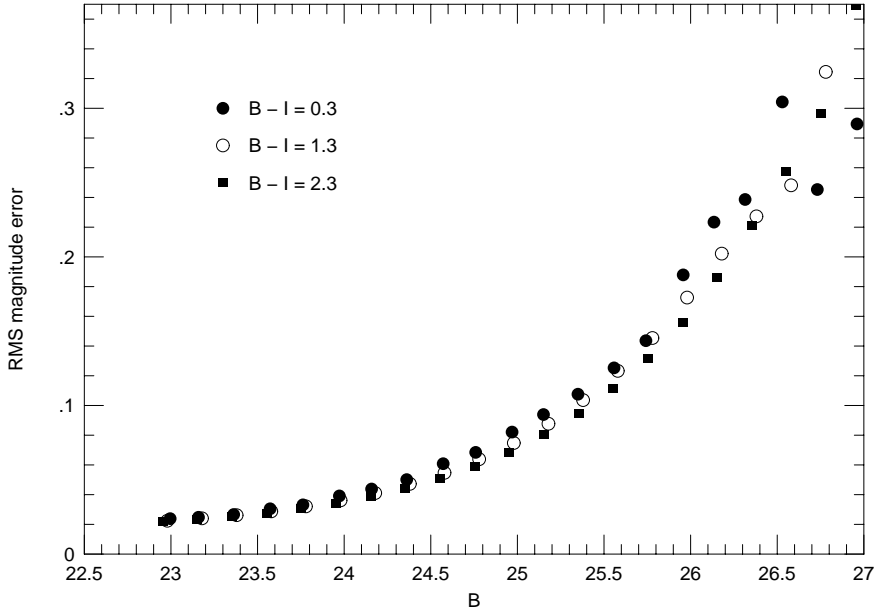}
\caption{Estimates of the RMS uncertainties in magnitudes for
clusters in NGC 1399, as derived from a comparison of the input
magnitudes of model clusters and the magnitudes returned after
processing. \label{fig:returned}}
\end{figure}

Completeness tests were carried out by adding a total of $\approx
3000$ artificial stars to each image and processing the results in a
manner identical to that used for the original data. Artificial
globulars (generated from a composite PSF derived from real images in
each frame) were added in batches of 100 with successively fainter $B$
magnitudes, and with colors $B - I = 0.3, 1.3,$ and 2.3 magnitudes,
respectively.  The results of these tests are shown for NGC 1399 in
Figure \ref{fig:compfig} and Figure \ref{fig:returned}. The 50\%
completeness level for the bulk of the detected objects occurs at $B
\approx 26.8$ in the WF frames, and $\approx 25.8$ in the PC images.
For F0336, the 50\% completeness level for the same color occurs at $B
\approx 27.3$ in the WF frames, and $B \approx 26.9$ in the PC.  The
completeness tests for NGC 1399 yield an estimate for the photometric
uncertainty for $1.3 < B - I < 2.3$ of $\approx 0.08$ mag rms at $B$ =
25, and $\approx 0.17$ mag rms at $B$ = 26.

\section{The Color Distributions. \label{sec:color}}

\begin{figure}[ht]
\epsfxsize=4.0in
\epsfbox[-96 72 504 700]{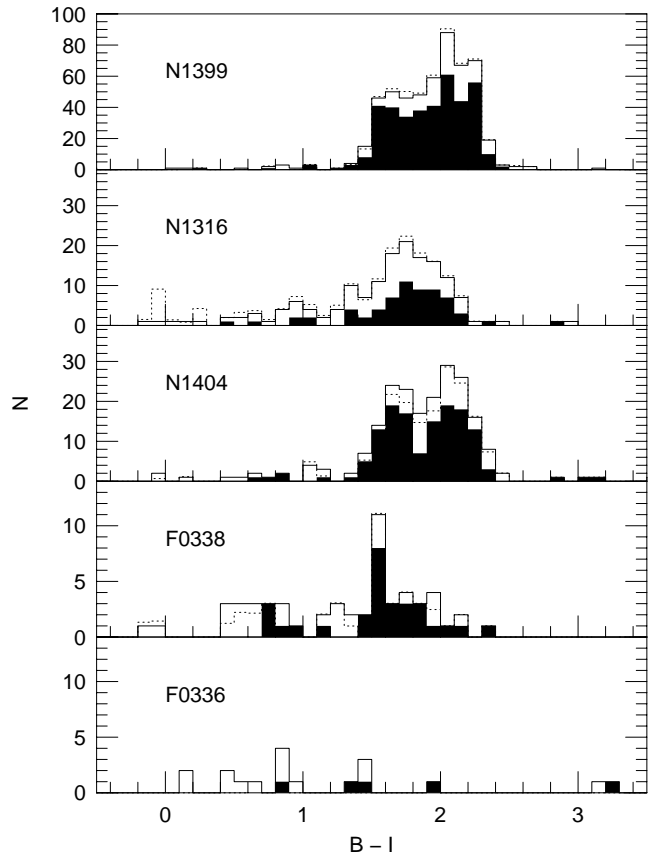}
\caption{Color distribution of compact sources in the five Fornax
fields. The shaded regions indicate the numbers of objects with $B <
25.0$, while the solid lines indicate the numbers of objects with $B <
26.0$. The dotted lines correspond to the completeness-corrected and
background-subtracted counts.  The dotted line shown for NGC 1404
shows the effect of background-subtraction using the color
distribution observed in the F0338 field and thereby better
characterizes the color distribution of globular clusters native to
NGC 1404. \label{fig:color}}
\end{figure}

In Figure \ref{fig:color} we show the numbers of globular clusters
found per 0.1 mag color interval brighter than $B = 26.0$ in each of
the five fields. The corresponding surface densities for the three
galaxies are tabulated in Table 2. The shaded regions in Figure
\ref{fig:color} show the color distributions for objects brighter than
$B = 25.0$. Assuming that the color distribution of any intergalactic
globular clusters roaming the Fornax cluster should be similar to that
seen in the three galaxies in our sample, the distribution apparent in
F0336 indicates that such clusters are rare, and that the apparent
distribution of colors is primarily representative of unresolved
background galaxies. Completeness fractions were computed by
spline-interpolation of our color-magnitude completeness grid.
Correcting each bin by simply dividing by the computed completeness
fraction and subtracting the F0336 distribution yields the
distributions indicated by the dotted lines in Figure \ref{fig:color}.

\begin{deluxetable}{cccccccc}
\tablecaption{Fornax Globular Cluster Color Distribution}
\tablecolumns{8}
\tablehead{
\colhead{} &
\multicolumn{2}{c}{NGC 1399} &
\multicolumn{3}{c}{NGC 1404} &
\multicolumn{2}{c}{NGC 1316} \\
\colhead{$B-I$} &
\colhead{N} &
\colhead{$f$\tablenotemark{a}} &
\colhead{N} &
\colhead{$f$\tablenotemark{a}} &
\colhead{$f$\tablenotemark{b}} &
\colhead{N} &
\colhead{$f$\tablenotemark{a}} \\
\colhead{} &
\colhead{} &
\colhead{(arcmin$^{-2}$)} &
\colhead{} &
\colhead{(arcmin$^{-2}$)} &
\colhead{(arcmin$^{-2}$)} &
\colhead{} &
\colhead{(arcmin$^{-2}$)} \\
}
\startdata
 -0.1--  0.0 &    0 &   0.0 &    2 &   0.4 &   0.1 &    1 &   3.8 \nl
  0.0--  0.1 &    1 &   0.0 &    0 &   0.0 &   0.0 &    1 &   0.6 \nl
  0.1--  0.2 &    1 &  -0.2 &    1 &  -0.2 &   0.2 &    1 &   0.3 \nl
  0.2--  0.3 &    1 &   0.2 &    0 &   0.0 &   0.0 &    1 &   1.7 \nl
  0.3--  0.4 &    0 &   0.0 &    0 &   0.0 &   0.0 &    0 &   0.0 \nl
  0.4--  0.5 &    0 &  -0.4 &    1 &  -0.2 &  -0.4 &    2 &   0.8 \nl
  0.5--  0.6 &    1 &   0.0 &    1 &   0.0 &  -0.4 &    2 &   1.4 \nl
  0.6--  0.7 &    0 &  -0.2 &    2 &   0.3 &  -0.1 &    3 &   1.5 \nl
  0.7--  0.8 &    2 &   0.5 &    1 &   0.2 &  -0.4 &    1 &   0.6 \nl
  0.8--  0.9 &    3 &  -0.1 &    2 &  -0.5 &  -0.2 &    4 &   1.7 \nl
  0.9--  1.0 &    1 &   0.0 &    0 &  -0.2 &  -0.2 &    6 &   3.0 \nl
  1.0--  1.1 &    3 &   0.7 &    4 &   1.0 &   1.0 &    4 &   2.2 \nl
  1.1--  1.2 &    0 &   0.0 &    3 &   0.7 &   0.3 &    2 &   1.0 \nl
  1.2--  1.3 &    1 &   0.2 &    0 &   0.0 &  -0.6 &    4 &   2.1 \nl
  1.3--  1.4 &    4 &   0.7 &    2 &   0.2 &   0.0 &   10 &   4.3 \nl
  1.4--  1.5 &   15 &   2.8 &    7 &   0.9 &   1.1 &    7 &   2.7 \nl
  1.5--  1.6 &   46 &   9.7 &   14 &   2.7 &   0.6 &   11 &   4.8 \nl
  1.6--  1.7 &   50 &  10.8 &   24 &   5.1 &   4.5 &   18 &   8.0 \nl
  1.7--  1.8 &   46 &  10.4 &   23 &   4.8 &   4.1 &   21 &   9.2 \nl
  1.8--  1.9 &   48 &  10.2 &   17 &   3.6 &   3.0 &   17 &   7.5 \nl
  1.9--  2.0 &   59 &  12.6 &   21 &   4.1 &   3.6 &   16 &   6.6 \nl
  2.0--  2.1 &   88 &  18.7 &   29 &   6.1 &   5.9 &   12 &   5.1 \nl
  2.1--  2.2 &   67 &  14.2 &   26 &   5.5 &   5.1 &    7 &   3.1 \nl
  2.2--  2.3 &   70 &  14.7 &   16 &   3.4 &   3.4 &    1 &   0.4 \nl
  2.3--  2.4 &   19 &   4.0 &    8 &   1.7 &   1.5 &    1 &   0.4 \nl
  2.4--  2.5 &    3 &   0.6 &    2 &   0.4 &   0.4 &    1 &   0.4 \nl
  2.5--  2.6 &    2 &   0.5 &    0 &   0.0 &   0.0 &    0 &   0.0 \nl
  2.6--  2.7 &    2 &   0.4 &    0 &   0.0 &   0.0 &    0 &   0.0 \nl
  2.7--  2.8 &    0 &   0.0 &    0 &   0.0 &   0.0 &    0 &   0.0 \nl
\enddata
\tablecomments{}
\tablenotetext{a}{Surface density after completeness correction and subtraction of the background as determined from F0336.}
\tablenotetext{b}{Surface density after completeness correction and subtraction of the color distribution in F0338, which lies on the opposite side of NGC 1399 at a distance similar to that of NGC 1404.}
\end{deluxetable}

The color distributions of globular clusters in both NGC 1399 and NGC
1404 are evidently bimodal. Moreover, the colors of the peaks and the
overall width of the color distributions are almost identical, though
NGC 1399 may have a slightly larger proportion of red clusters. The
mean $B - I$ colors of the two samples are $2.01 \pm 0.46$ and $1.98
\pm 0.58$, and a Kolmogorov-Smirnov test indicates that the hypothesis
that the two populations were drawn from the same parent population
cannot be rejected at the 63\% confidence level.  Based on the ratios
of the numbers of globular clusters counted to $B = 25$ and $B = 26$,
it appears that the bluer clusters are on average somewhat brighter
than the red clusters. We discuss this finding further in Section
\ref{sec:lf} and in Forbes et al. (1998).

The higher surface density of globular clusters around NGC 1404, as
well as the difference in mean colors when compared with a similar
field on the opposite side of NGC 1399 (F0338), indicate that these
globular clusters indeed belong to NGC 1404 as opposed to the much
more populous and extended population of NGC 1399.  Taking into
account the orientations and the area-weighted centers of the WFPC2
field-of-view in NGC 1404 and F0338, we find that the F0338 field is
0.39\arcmin~ closer in projection to NGC 1399 than the NGC 1404 field.
Assuming a radial surface density profile of NGC 1399 globular
clusters which goes as $f \propto r^{-1.5}$ (\eg Kissler-Patig et al.
1997a), we apply a correction factor of 0.93 to the surface density of
globulars detected in F0338 to match the surface density of NGC 1399
globulars expected at the radius of NGC 1404. Subtracting the
resulting, completeness-corrected distribution of globular clusters in
F0338 from that of NGC 1404 yields the dotted line in Figure
\ref{fig:color}.  The total number of globular clusters and the mean
color are minimally affected, supporting the notion that the majority
of these globular clusters are indigenous to NGC 1404. Furthermore,
the similarity in the color distributions of NGC 1399 and NGC 1404 are
consistent with the idea that NGC 1399's overabundance of globular
clusters may have come partly at the expense of NGC 1404 and other
galaxies in the Fornax cluster (Forbes et al. 1997b; Kissler-Patig et
al. 1998). This does not, of course, rule out the possibility that the
red and blue clusters originated in different galaxies or under
different circumstances.

Interestingly, the color distribution of compact sources in F0338 appears to
be significantly different from those seen in the galaxy-centered
fields.  Even after subtraction of the color distribution in F0336
(the background field), the colors of objects in F0338 peak at $B - I
= 0.6$, which is 0.1 magnitudes bluer than the centroid of the bluest
peak in either NGC 1399 or NGC 1404.  There is also a group of objects
with $B - I < 1$ which appears to significantly exceed the number of
such objects found in the F0336.  These findings are consistent with
the presence of a significant color gradient, as first reported by
Ostrov et al.  (1993). These observations and their consequences are
discussed in detail by Forbes et al. (1998).

In spite of the possibility of unseen, distributed dust, the
``unobscured'' objects in NGC 1316 are on average 0.45 magnitudes
bluer in $B - I$ than either those in NGC 1399 or NGC 1404. The peak
in the number counts occurs just redward of the position of the bluest
peak in NGC 1399 and NGC 1404, with a significant tail of objects with
$B - I < 1.2$. The bluer objects are much less luminous than the
bright, blue clusters found by Holtzman et al. (1992) in NGC 1275. The
lack of bright, blue globulars, while perhaps surprising given the
ample evidence for a recent merger event, agrees with the findings of
Shaya et al. (1996).  However, this may not be entirely surprising
given the much smaller quantity of dust visible in NGC 1316 than in
NGC 1275.  And despite the arguments of Shaya et al. (1996), we cannot
rule out bright young clusters in the obscured, highly reddened
central regions of the galaxy.  Very young ($\tau \sim$ 5 Myrs) {\it
stellar} populations do produce stars as bright as $M_B = -9$.
However, such stars would mostly be far hotter and bluer than the
objects we detect in NGC 1316. Shown in Figure \ref{fig:cm} is the
$B-I$ color-magnitude sequence of the turnoff and horizontal branch
expected for a stellar population of age $4 \times 10^7$ yrs and
[Fe/H] = -0.4 (Worthey, private communication). While bearing a
general resemblance to the distribution of the faintest objects in NGC
1316, the luminosity function for such a stellar population requires a
very much larger number of hot, blue objects ($B-I < 0.5$) than we
actually find.

 Many of the objects surrounding NGC 1316 are similar in color and
magnitude to the old open clusters or the faint, metal-poor, outlying
globular clusters in our own Galaxy such as PAL 1 and Eridanus. Based
on the models of Worthey (1994), the objects with $25 < B < 26$ and
$1.2< B - I < 2.2$ are consistent with a population of old, metal-poor
clusters with masses of order $10^5$ M$_{\odot}$, or metal-rich
clusters with ages as young as 1 Gyr and masses of order $10^4$
M$_{\odot}$. We discuss the nature of these objects further in Section
\ref{sec:n1316}.

Both Whitmore et al. (1995) and Elson \& Santiago (1996a, 1996b) used
{\it HST} data to confirm the presence of color dichotomies in the GC
system of M 87 first suggested by the ground-based data of Lee \&
Geisler (1993).  The color differences between the blue and red peaks
in the color histograms of globular clusters in M 87 are $\Delta ( V -
I )_0 \approx 0.25$ magnitudes. However, the $B-I$ color is about
twice as sensitive to metallicity as $V-I$. If we assume that age is
not an important contributor to the spread in color, and if we use
Couture, Harris, \& Allwright's (1990) color-metallicity calibrations
for Milky Way globular clusters, we find that the
full-width-at-half-maxima for the color distributions of Elson \&
Santiago (1996) and Whitmore et al.  (1995) ($\Delta ( V - I )_0
\approx 0.45$ mag) correspond to a spread in [Fe/H] of about 2.3 dex.
NGC 1399 and NGC 1404 have $\Delta ( B - I )_0 \approx 0.8$ in both
NGC 1399 and NGC 1404, yielding a metallicity spread $\Delta$ [Fe/H]
$\approx 2.1$ dex.  These values are in good agreement with the
metallicity spread inferred from $C - T_1$ measurements of the
brighter globulars in NGC 1399 by Ostrov et al.  (1993). Using only
the color differences between peaks in M 87, NGC 1399, and NGC 1404,
we find $\Delta$ [Fe/H] $\approx 1.3, 1.1,$ and 1.1 dex, respectively.
As found in previous ground-based studies (see Geisler et al.  1996
for a tabulation), the ranges of globular cluster metallicities to be
found in M 87 and NGC 1399 (and other bright ellipticals) are similar,
spanning an interval which extends from the metal-poorest globulars in
our Galaxy to a metallicity significantly higher than that of the sun.

\clearpage

\section{The Luminosity Functions. \label{sec:lf}}

\begin{figure}[ht]
\epsfxsize=4.0in
\epsfbox[-160 72 440 710]{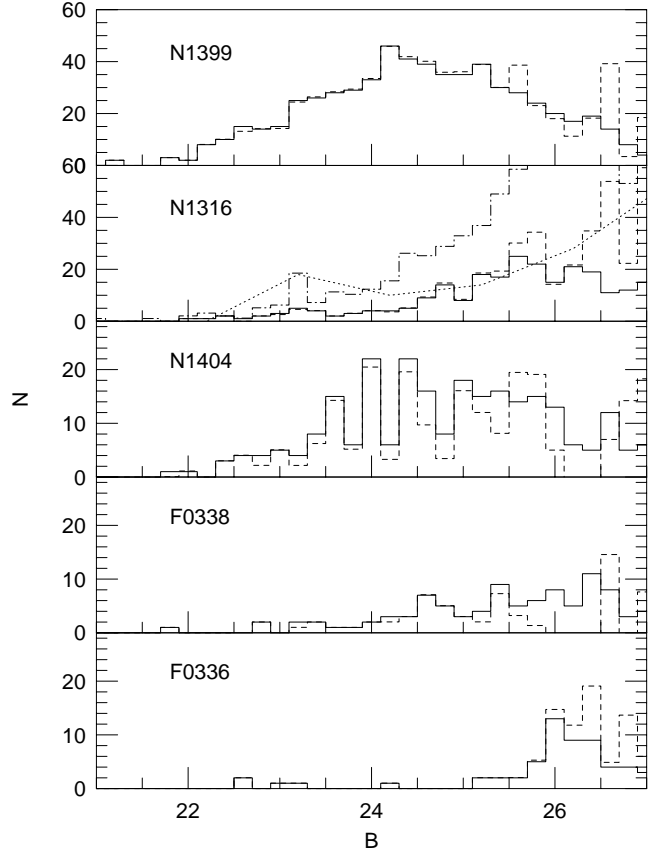}
\caption{Luminosity functions of point sources in the five Fornax
fields. For NGC 1399, NGC 1316, and F0338, the dashed lines show the
effects of dividing by the computed completeness fractions and
subtracting the completeness-corrected luminosity function observed in
field F0336. For NGC 1404, the dashed curve shows the effect of
subtracting the F0338 luminosity function from observed distribution
in NGC 1404.  The dashed line in the F0336 panel simply shows the
effect of dividing by the computed completeness fractions. The
dash-dot lines for NGC 1316 shows the effect of including those 
objects detected in the obscured regions in the WF chips. The dotted
curve for NGC 1316 shows the ``completeness-corrected'' luminosity
function of Galactic open clusters tabulated by van den Bergh \&
Lafontaine (1984), multiplied by a factor of 10.  \label{fig:lf}}
\end{figure}

\begin{figure}[ht]
\epsfxsize=4.0in
\epsfbox[-160 72 440 690]{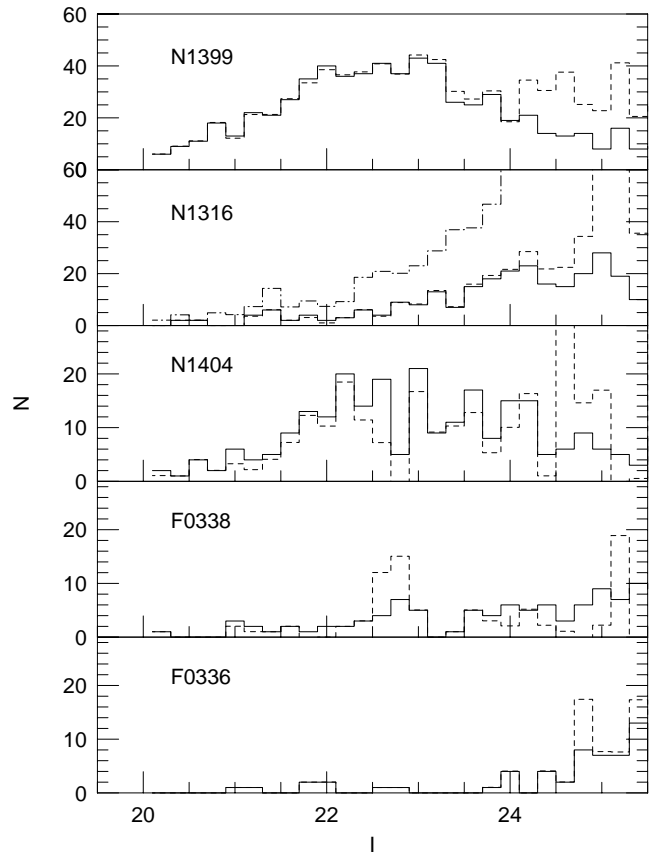}
\caption{$I$-band luminosity functions of point sources in the five
Fornax fields. The dashed lines are as described in Figure 7.
\label{fig:lfi}}
\end{figure}

The luminosity functions (LFs) for our three target galaxies are given
in Table 3 and are plotted in Figures \ref{fig:lf} and
\ref{fig:lfi}. The plotted LFs for NGC 1399 and NGC 1316 have been
corrected for completeness, and background-corrected by subtracting
the observed, completeness-corrected LF of F0336.  Similarly, the GCLF
for NGC 1404 has been adjusted using the completeness-corrected (but
not background subtracted) LF in F0338 after scaling by a factor of
0.93 as described earlier to account for the fact that the center of
the F0338 field is 0\Min4 closer to NGC 1399 than is NGC
1404. Completeness corrections were made for each globular cluster
counted based on the WFPC2 chip on which it was found, but no account
was taken of the cluster's position on the chip.

\begin{deluxetable}{ccccccc}
\tablecaption{Fornax Globular Cluster Luminosity Functions}
\tablecolumns{7}
\tablehead{
\colhead{} &
\multicolumn{2}{c}{NGC 1399} &
\multicolumn{2}{c}{NGC 1404} &
\multicolumn{2}{c}{NGC 1316} \\
\colhead{$B$} &
\colhead{N} &
\colhead{$f$\tablenotemark{a}} &
\colhead{N} &
\colhead{$f$} &
\colhead{N} &
\colhead{$f$} \\
\colhead{} &
\colhead{} &
\colhead{(arcmin$^{-2}$)} &
\colhead{} &
\colhead{(arcmin$^{-2}$)} &
\colhead{} &
\colhead{(arcmin$^{-2}$)} \\
}
\startdata
 20.1--20.3 &    0 &   0.0 &    0 &   0.0 &    0 &   0.0 \nl
 20.3--20.5 &    0 &   0.0 &    0 &   0.0 &    1 &   0.4 \nl
 20.5--20.7 &    0 &   0.0 &    1 &   0.0 &    0 &   0.0 \nl
 20.7--20.9 &    0 &   0.0 &    1 &   0.2 &    0 &   0.0 \nl
 20.9--21.1 &    0 &   0.0 &    0 &   0.0 &    0 &   0.0 \nl
 21.1--21.3 &    2 &   0.4 &    0 &   0.0 &    0 &   0.0 \nl
 21.3--21.5 &    0 &   0.0 &    0 &   0.0 &    0 &   0.0 \nl
 21.5--21.7 &    0 &   0.0 &    0 &   0.0 &    0 &   0.0 \nl
 21.7--21.9 &    3 &   0.6 &    1 &   0.0 &    0 &   0.0 \nl
 21.9--22.1 &    2 &   0.4 &    1 &   0.2 &    1 &   0.4 \nl
 22.1--22.3 &    8 &   1.7 &    0 &   0.0 &    1 &   0.4 \nl
 22.3--22.5 &   10 &   1.9 &    3 &   0.6 &    2 &   0.6 \nl
 22.5--22.7 &   15 &   2.5 &    4 &   0.4 &    1 &  -0.2 \nl
 22.7--22.9 &   14 &   2.9 &    4 &   0.8 &    2 &   0.8 \nl
 22.9--23.1 &   15 &   3.2 &    5 &   0.7 &    3 &   1.3 \nl
 23.1--23.3 &   25 &   5.3 &    4 &   0.4 &    5 &   2.1 \nl
 23.3--23.5 &   26 &   5.5 &    8 &   1.3 &    4 &   1.7 \nl
 23.5--23.7 &   28 &   5.9 &   15 &   2.9 &    2 &   0.8 \nl
 23.7--23.9 &   29 &   6.1 &    6 &   0.9 &    3 &   1.3 \nl
 23.9--24.1 &   33 &   6.7 &   22 &   4.2 &    4 &   1.5 \nl
 24.1--24.3 &   46 &   9.7 &    6 &   1.1 &    4 &   1.7 \nl
 24.3--24.5 &   41 &   8.7 &   22 &   3.5 &    5 &   2.1 \nl
 24.5--24.7 &   39 &   8.3 &   16 &   2.0 &    9 &   3.8 \nl
 24.7--24.9 &   35 &   7.2 &    8 &   0.7 &   14 &   5.9 \nl
 24.9--25.1 &   35 &   7.5 &   18 &   3.5 &    8 &   3.5 \nl
 25.1--25.3 &   39 &   7.8 &   15 &   1.9 &   18 &   7.5 \nl
 25.3--25.5 &   30 &   6.4 &   16 &   2.3 &   17 &   8.2 \nl
 25.5--25.7 &   28 &   7.4 &   14 &   3.8 &   25 &  11.8 \nl
 25.7--25.9 &   24 &   4.5 &   15 &   3.7 &   22 &  13.9 \nl
 25.9--26.1 &   20 &   4.0 &   13 &   1.3 &   15 &   6.1 \nl
 26.1--26.3 &   17 &   1.6 &    6 &   0.1 &   21 &   8.2 \nl
 26.3--26.5 &   19 &   3.9 &    5 &  -2.0 &   19 &  14.5 \nl
\enddata
\tablecomments{}
\tablenotetext{a}{Surface density after background subtraction and completeness correction.}
\end{deluxetable}

The LF for objects in NGC 1316 is clearly different
from those of NGC 1399 and NGC 1404. In addition to a bluer mean
color, the number of objects in NGC 1316 increases rapidly faintwards
and does not show a log-normal distribution of the kind normally seen
in globular cluster systems.  We discuss NGC 1316's cluster system
in more detail below.

The GCLFs for NGC 1399 and NGC 1404 appear to be very similar. A
straightforward comparison using the Kolmogorov-Smirnov test reveals
that the null hypothesis (that the two populations were drawn from the
same parent population) cannot be rejected at the 90\% confidence
level. If the magnitudes of NGC 1404 globular clusters are offset by
$B = -0.27$ magnitudes (see below), the K-S test yields a probability of 
74\% in favor of the null hypothesis.  After binning, completeness
correction, and background subtraction, a $\chi^2$ test over the range
$20 < B < 26.5$ gives an 86\% probability in favor of the null hypothesis.

\begin{figure}[ht]
\epsfxsize=3.0in
\epsfbox[-250 72 350 700]{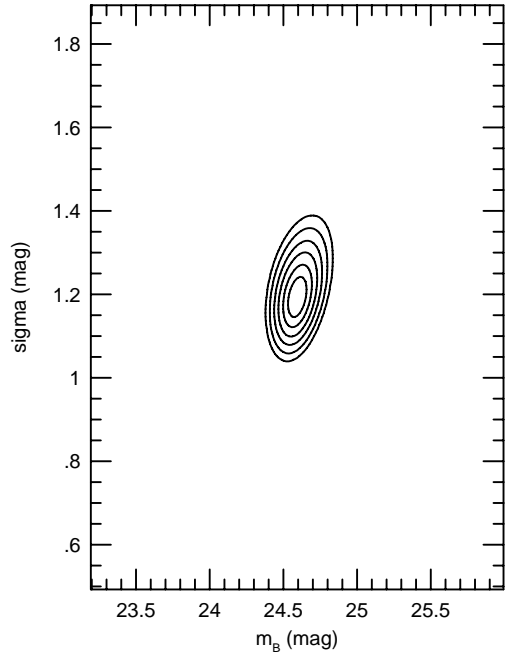}
\caption{Contours of $\chi^2$ in the $\langle m_V^0
\rangle$-$\sigma$ plane for the combined luminosity functions of NGC
1399 and NGC 1404, using the method of Secker \& Harris (1993). The
contours range from $0.5\sigma$ to $3\sigma$ and are spaced at
intervals of $0.5\sigma$. \label{fig:contours}}

\end{figure}

\begin{figure}[ht]
\epsfxsize=4.0in
\epsfbox[-100 72 500 680]{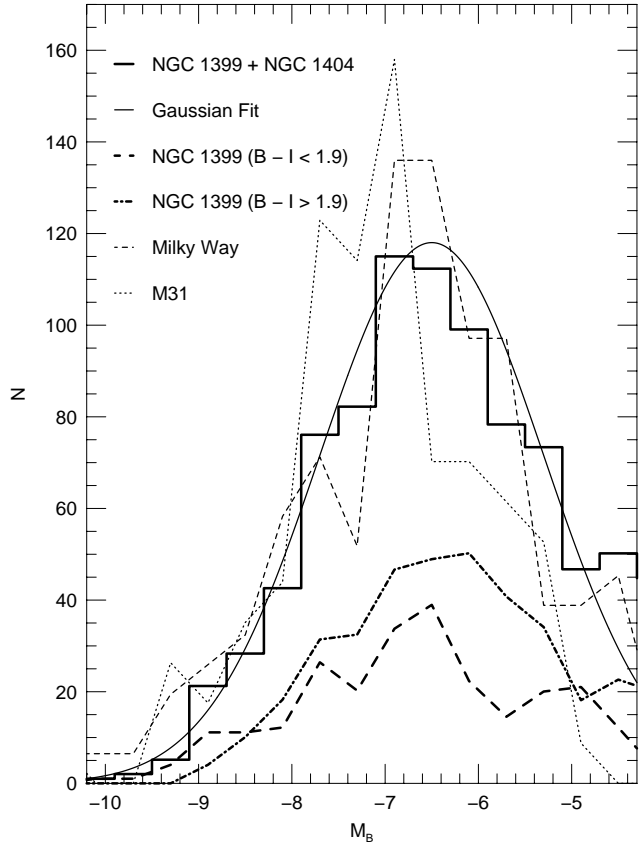}
\caption{Luminosity functions of NGC 1399 and NGC 1404 compared
with the luminosity functions for halo globulars in the Galaxy and
M 31. The Local Group luminosity functions have been normalized to the
same total number of globulars as found in the combined NGC 1399 and
NGC 1404 samples. Also shown are separate luminosity functions for the
blue and red globulars in NGC 1399. The Gaussian fit results from the
application of version 2 of Secker \& Harris's (1993)
maximum-likelihood estimator. The difference between the apparent peak
in the uncorrected data and that of the Gaussian are due to
incompleteness and photometric errors, which are properly handled by
the program. \label{fig:lg}}
\end{figure}

We have used version 2.0 of the peak-finding code described by Secker
\& Harris (1993) (and kindly provided by J. Secker) to compute the
maximum-likelihood peaks and dispersions of both Gaussian and
Student's $t_5$ functions from our GCLF sample. Owing to the current
limitations of the code we have used only the globular clusters
detected in the WF frames, for which the completeness functions are
reasonably similar. As we are neglecting only 10\% (20\%) of the total
number of clusters detected around NGC 1399 (NGC 1404), the statistics
are not seriously affected. For the present purposes we use only 
clusters with colors in the range $1.2 < B - I < 2.6$. The results are
given in Table 4.  We also include maximum-likelihood fits to the
Milky Way and M 31 cluster data described below, and we quote the
values given by Whitmore et al.  (1995) for the {\it HST} $V$-band
data of M 87. We assume reddenings of E(B-V) = 0.11 and 0.01 for M 31
and Fornax, respectively (Burstein \& Heiles 1982; Schlegel,
Finkbeiner, \& Davis 1998).  Figure \ref{fig:contours} shows the
$\chi^2$ contours for the NGC 1399 $B$-band GCLF Gaussian
fit in the peak-dispersion plane.  For each entry in Table 4 the
listed uncertainties correspond to the projections of the $1\sigma$
contours onto the peak magnitude and dispersion axes.  The Gaussian
which best fits the combined NGC 1399 + NGC 1404 data is shown in
Figure \ref{fig:lg}. The value of $\langle m_V^0 \rangle = 23.73 \pm
0.06$ we find for NGC 1399 is in good agreement with
the value of $\langle m_V^0 \rangle = 23.77 \pm 0.06$ (after
extinction correction using our adopted reddening) determined by Kohle
et al.  (1996) from ground-based data.

\clearpage

\begin{deluxetable}{lcccccc}
\tablecaption{Parametric Fits to Globular Cluster Luminosity Functions}
\tablecolumns{7}
\tablehead{
\colhead{} &
\colhead{$\langle m_B^0\rangle $} &
\colhead{$\sigma$} &
\colhead{$\langle m_B^0 \rangle $} &
\colhead{$\sigma$} &
\colhead{$\langle m_V^0\rangle $} &
\colhead{$\sigma$} \\
\colhead{} &
\multicolumn{2}{c}{(Gaussian)} &
\multicolumn{2}{c}{($t_5$)} &
\multicolumn{2}{c}{(Gaussian)} \\
}
\startdata
Milky Way & -6.49 $\pm$ 0.19 & 1.54 $\pm$ 0.15 & -6.61 $\pm$ 0.17 & 1.28 $\pm$ 0.15 &  &  \nl
M 31 & 17.38 $\pm$ 0.11 & 1.05 $\pm$ 0.08 & 17.41 $\pm$ 0.11 & 0.90 $\pm$ 0.09 &  &  \nl
NGC 1399 & 24.55 $\pm$ 0.08 & 1.19 $\pm$ 0.05 & 24.56 $\pm$ 0.08  & 1.07 $\pm$ 0.06 & 23.73 $\pm$ 0.08  & 1.19 $\pm$ 0.05 \nl
NGC 1404 & 24.82 $\pm$ 0.20  & 1.41 $\pm$ 0.15 & 24.84 $\pm$ 0.21  & 1.20 $\pm$ 0.15  &  &  \nl
NGC 1399 ($B-I > 1.9$) & 24.71 $\pm$ 0.08 & 1.16 $\pm$ 0.06 & 24.70 $\pm$ 0.08 & 1.05 $\pm$ 0.07 &  &  \nl
NGC 1399 ($B-I < 1.9$) & 24.32 $\pm$ 0.12 & 1.24 $\pm$ 0.09  & 24.34 $\pm$ 0.09 & 1.10 $\pm$ 0.08 &  &  \nl
NGC 1399 + NGC 1404 & 24.62 $\pm$ 0.07  & 1.23 $\pm$ 0.05 & 24.62 $\pm$ 0.08 & 1.09 $\pm$ 0.05 & 23.77 $\pm$ 0.06 & 1.22 $\pm$ 0.05 \nl
M 87\tablenotemark{a}  & &  &  &  & 23.72 $\pm$ 0.06 & 1.40 $\pm$ 0.06 \nl
\enddata

\tablecomments{}
\tablenotetext{a}{Values quoted from Whitmore et al. 1995.}

\end{deluxetable}

The peak magnitudes computed for NGC 1399 and NGC 1404 differ by $0.27
\pm 0.22$ magnitudes in the sense that NGC 1404 may lie some 2.1 Mpc
beyond NGC 1399, though the uncertainty is large. The GCLF width for
NGC 1404 is evidently slightly broader than that for NGC 1399, though
again the uncertainties on the NGC 1404 result do not rule out
identical widths.

As this is the first time we have been able to examine the LF of
Fornax globular clusters to this depth, it is interesting to compare
it with the LFs of Local Group globular clusters.  In Figure
\ref{fig:lg} we compare the background-corrected, combined GCLF for
NGC 1399 and NGC 1404 (after shifting the latter by $-0.27$ magnitudes)
to the LFs of globular clusters in our own Galaxy and M 31.  The Milky
Way GCLF derives from the halo globular cluster compilation of Harris
(1996) on the McMaster University WWW
page\footnote[2]{http://www.physics.mcmaster.ca/Globular.html}. The
data for globular clusters in M 31 comes from the halo sample
tabulated by Reed et al. (1994), and we assume (m--M)$_0$ = 24.43 for
M 31 (Ajhar et al. 1995).  The Local Group GCLFs are compared with the
completeness- and background-corrected GCLF of NGC 1399 + NGC 1404,
shifted assumed (m--M)$_0$ = 31.05 (see below) and $E(B-V) =
0.01$ (Schlegel, Finkbeiner, \& Davis 1998). The Local Group GCLFs
have been normalized to the total number of GCs counted in NGC 1399
and NGC 1404 with $M_B < -5$.

While the Milky Way GCLF appears in most respects to be fairly similar
in form to the GCLF for NGC 1399 + 1404, the GCLF for M 31 globular
clusters appears to be somewhat more peaked. The computed GCLF widths
determined for the Galaxy and M 31 are broader and narrower,
respectively, than we find for the early-type galaxies. The
uncertainty on the GCLF width of Galactic globulars is large enough to
accommodate the Fornax and M 87 results, but the M 31 distribution is
clearly at odds with all others. The mean magnitudes and widths we
find for the M 31 GCLF are consistent with those determined for the
outer-halo M 31 globulars by Kavelaars \& Hanes (1997). Incompleteness
of the M 31 sample at the faint end, though difficult to quantify,
must be at least partially responsible for this discrepancy.  In any
event, the numbers of Local Group globular clusters are too small to
allow strong conclusions to be made. A $\chi^2$ test comparing the NGC
1399 + 1404 GCLF with the Milky Way and M 31 GCLFs indicates that the
null hypothesis remains valid at the 66\% and 63\% confidence levels,
respectively. Similarly comparing the NGC 1399 + 1404 GCLF with the
combined Milky Way and M 31 GCLFs yields a probability for the null
hypothesis of greater than 64\%.

Ashman, Conti, \& Zepf (1995) have shown that good agreement between
the measured GCLF peaks of early-type and late-type galaxies could be
achieved by accounting for the different opacities and increased
line-blanketing expected in higher metallicity systems.  For NGC 1399
in particular, they predict an offset $\Delta \langle m_B^0 \rangle =
0.26$ mag with respect to the peak observed in the Milky Way. Using
this offset and the values in Table 4, we find $(m-M)_0 = 30.78 \pm
0.21$. Ashman et al.'s (1995) models were computed using only Gaussian
dispersions.  However, since their model distributions were symmetric,
the $\Delta \langle m_B^0 \rangle$ for $t_5$ distribution should be
very similar. The peak magnitudes computed for the Milky Way and NGC
1399 using the $t_5$ function then yield $(m-M)_0$ = 30.91 $\pm$ 0.19.
The agreement between these distance moduli and those determined using
other distance-measuring techniques (see below) underscores the
remarkable uniformity in the underlying mass functions of globular
clusters in very different galaxy types.

Also shown in Figure \ref{fig:lg} are the LFs for the 230 bluer ($B-I
< 1.9$) and the 348 redder ($B-I > 1.9$) globular clusters in NGC
1399.  According to Table 4 there is no significant difference between
the widths of these two distributions, and it appears from Figure
\ref{fig:lg} that a simple faintward shift of the blue distribution
could bring it into line with the red one. This is reflected in the
mean and peak magnitudes in Table 4, which differ by $0.36 \pm 0.12$
mags ($t_5$) between the red and blue samples. After shifting the red
LF by -0.36 magnitudes, a $\chi^2$ test gives a probability for the null
hypothesis of 51\%.  Current data for two positions in M 87 (Whitmore
et al.  1995; Elson \& Santiago 1996b) and NGC 5846 (Forbes et al.
1997a) are consistent with the blue peak being from 0.0 to 0.3
magnitudes brighter in $V$ than the red peak. If attributed entirely
to metallicity effects, the 0.36 $B$-band magnitude difference we find
for NGC 1399 clusters would correspond to $\Delta [Fe/H] \approx 1$
dex (Couture et al. 1990), and would produce a $V$-band difference
between the two peaks of $\approx 0.2$ magnitudes.

\begin{figure}[ht]
\epsfxsize=4.0in
\epsfbox[-160 72 428 736]{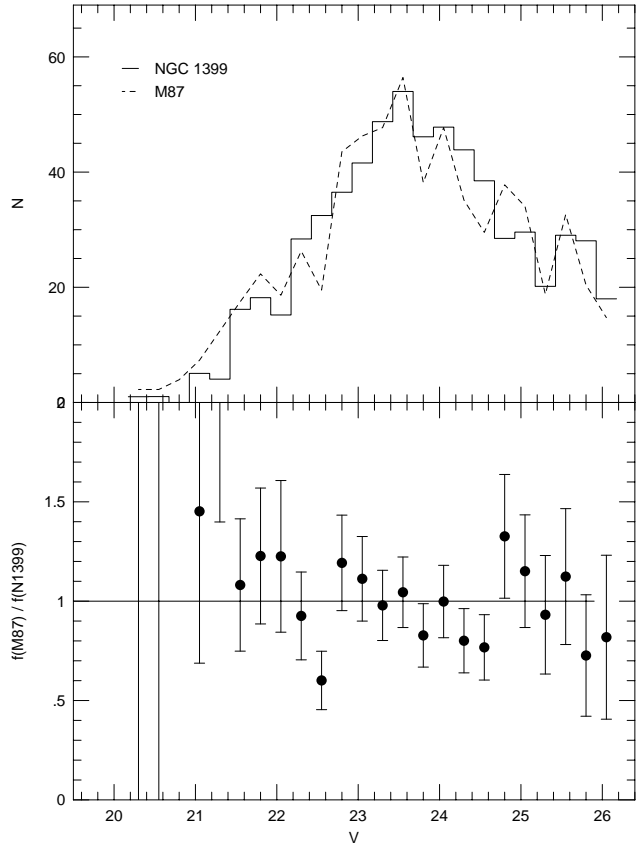}
\caption{The globular cluster luminosity function of NGC 1399,
after conversion to $V$-band magnitudes (as described in the text) and
shifting by $-0.1$ mag, compared with the {\it HST} GCLF determined by
Whitmore et al. (1995) for M 87.  The lower panel shows the bin-by-bin
ratio of the M 87 GCLF to the normalized GCLF of NGC 1399. The error
bars reflect Poisson statistics only.  \label{fig:m87}}
\end{figure}

In Figure \ref{fig:m87} we compare the GCLF for NGC 1399 
with the $V$-band, {\it HST} GCLF for M 87 of Whitmore et al. (1995).
We have converted our $B$-band measurements to $V$-band for the
comparison using

\begin{equation}
( B - V )_0 = 0.347 ( B - I )_0 + 0.163,
\end{equation}

\noindent which is a least-squares fit to the least-reddened (E(B-V)
$< 0.4$) Galactic globular clusters as compiled on the McMaster WWW
page.  Simply comparing values of $\langle m_V^0 \rangle$ in Table 4
reveals that the peak of the NGC 1399 GCLF is 0.01 $\pm$
0.10 magnitudes fainter than that of M 87. This is in good agreement
with offsets of $0.05 \pm 0.09$ magnitudes found by Kohle et al.
(1996) and $0.08 \pm 0.16$ magnitudes found by Blakeslee \& Tonry
(1996)) from deep, ground-based GCLF studies, after correction for our
adopted reddening. The NGC 1399 GCLF was recomputed after
applying a shift of $-0.01$ to the computed $V$ magnitudes to bring
the peak into line with that of M 87 as per Table 4. The resulting
$V$-band luminosity function is compared with a scaled version of
Whitmore et al.'s (1995) M 87 GCLF in the upper panel of Figure
\ref{fig:m87}.

There is clearly very little difference between the GCLFs of the two
cD galaxies. In the lower panel of Figure \ref{fig:m87} we show the
bin-by-bin ratio of the normalized M 87 luminosity function to that of
NGC 1399.  The GCLFs are evidently indistinguishable over most of the
range of comparison, though there may be proportionately more very
bright globulars in M 87 than there are in NGC 1399. A $\chi^2$ test
indicates that the null hypothesis cannot be rejected at the 69\%
confidence level.

The GCLF width listed in Table 4 for NGC 1399 is significantly
narrower ($\approx 4\sigma$) than that found by Whitmore et al. 1995
for M 87. This is due in part to the very bright M 87 globulars noted
above, and in part to differences in the methods used to measure the
widths; the method of Secker \& Harris (1993) we have adopted here
takes into account the broadening of the LF due to photometric errors.
The GCLF width we find for NGC 1399 is very similar to those found for
NGC 4278, NGC 4494 (Forbes 1996b) and NGC 1404, but is significantly
narrower than those of either NGC 4365 GCLF (Forbes 1996a) or NGC 5846
(Forbes et al. 1996), all measured using the code of Secker \& Harris
(1993).  Thus, NGC 1399 does not support Kissler-Patig et al.'s
(1997a) suggestion that GCLF properties vary little among the lesser
galaxies, be they early or late-type, but that they vary significantly
among cluster-dominating ellipticals.

Adopting for M 87 a distance of 16.1 $\pm$ 1.3 Mpc (Ferrarese et al.
1997; Yasuda et al. 1997), then the relative shift of 0.01 $\pm$ 0.1
magnitudes we find for the GCLF of NGC 1399 gives a distance to Fornax
of 16.2 $\pm$ 1.5 Mpc. This is in good agreement with a Fornax
distance of 16.9 $\pm$ 1.1 Mpc determined by McMillan, Ciardullo, \&
Jacoby (1993) using planetary nebula luminosity functions, though in
somewhat poorer agreement with the distance of 18.4 $\pm$ 1.8 Mpc
determined by Madore et al. (1996) using {\it HST} observations of
Cepheids in NGC 1365. While Cepheids are believed to be more accurate
distance indicators than GCLFs, it may also be that NGC 1365 lies well
out at the periphery of the Fornax cluster, as spiral galaxies are
wont to do. Since NGC 1399 sits almost exactly at the center of the
cluster's potential well, we can use our inferred distance to compute
a value for the Hubble constant without incurring the additional
uncertainty introduced by the expanse of the Fornax cluster itself. We
use a heliocentric velocity for Fornax of 1450 $\pm$ 34 km s$^{-1}$
(Held \& Mould 1994).  Correcting for the solar motion with respect to
the Local Group (--91 km s$^{-1}$; Yahil, Tammann, \& Sandage 1977),
and a Virgocentric infall component along the line of sight to Fornax
of 36 km s$^{-1}$ (Tammann \& Sandage 1985), we obtain an
expansion-induced velocity of 1323 $\pm 34$ km s$^{-1}$.  This then
yields a Hubble constant $H_0 = 82 \pm 8 $ km s$^{-1}$ Mpc$^{-1}$,
where the uncertainty reflects random measurement errors only. This
agrees reasonably well with a value of 73 $\pm$ 6 (random) $\pm$ 8
(systematic) km s$^{-1}$ Mpc$^{-1}$ found by Madore et al.  (1996)
using, among other things, {\it HST}-measured Cepheid distances to
several nearby galaxy groups including Virgo and Fornax.

\clearpage

\section{The Case of NGC 1316 \label{sec:n1316}}

Why is the LF of clusters in NGC 1316 not log-normal in form, as are
those of almost all globular cluster systems studied to date?  One
might postulate that the LF of objects surrounding NGC 1316 reflects a
young system which has yet to undergo depletion by the disruption of
its faintest members. Removal of the majority of objects fainter than
$B \approx 24.5$ could conceivably leave behind a log-normal LF.
However, the number of objects at $B = 24.5$ is small, and such a
depletion would leave NGC 1316 with roughly as many clusters as NGC
1404. The specific frequency of globular clusters in NGC 1404 is
already quite low ($S_N \sim 2$, Forbes et al. 1997), and given the
1.4 magnitude difference in total luminosity between these two
galaxies, we would be left with a remarkably low $S_N \approx 0.5$ for
NGC 1316. While this would be highly unusual for any early-type
galaxy, it would be even more surprising in view of NGC 1316's evident
merger history and apparent dominance of its corner of the Fornax
cluster. 

If NGC 1316 has undergone a significant starburst and is destined to
fade by at least a factor of four in the time required to disrupt the
fainter clusters, it is conceivable that $S_N$ might eventually
approach a value more typical of early-type galaxies.  However, a
direct comparison between the WF3 images of NGC 1316 and NGC 1404
reveals that the ratio in the total luminosities of these two galaxies
is largely preserved in surface brightness measurements at the larger
radii where NGC 1316 appears to be free of dust. It seems unlikely
that significant recent star formation has taken place at these large
radii, and that we must therefore be looking primarily at old stars.
Thus, even if there is substantial fading of a younger stellar
population residing at the center of NGC 1316, the total magnitude
(and therefore $S_N$) will be largely unaffected. 

The luminosity function of the objects we see in NGC 1316 has been
seen in other galaxies as well, NGC 4038/4039 (Whitmore \& Schweizer
1995) and NGC 3597 (Carlson et al. 1998) being two well-studied cases.
Superposed on the NGC 1316 luminosity function in Figure \ref{fig:lf}
is the ``completeness-corrected'' luminosity function of Galactic open
clusters tabulated by van den Bergh \& LaFontaine (1984), scaled
upwards by a factor of 10. While there may be relatively fewer bright
objects in NGC 1316 than there are open clusters in the Galaxy, the
luminosity functions are reasonably similar over the range for which
we have good statistics.

\begin{figure}[ht]
\epsfxsize=4.0in
\epsfbox[-72 36 528 536]{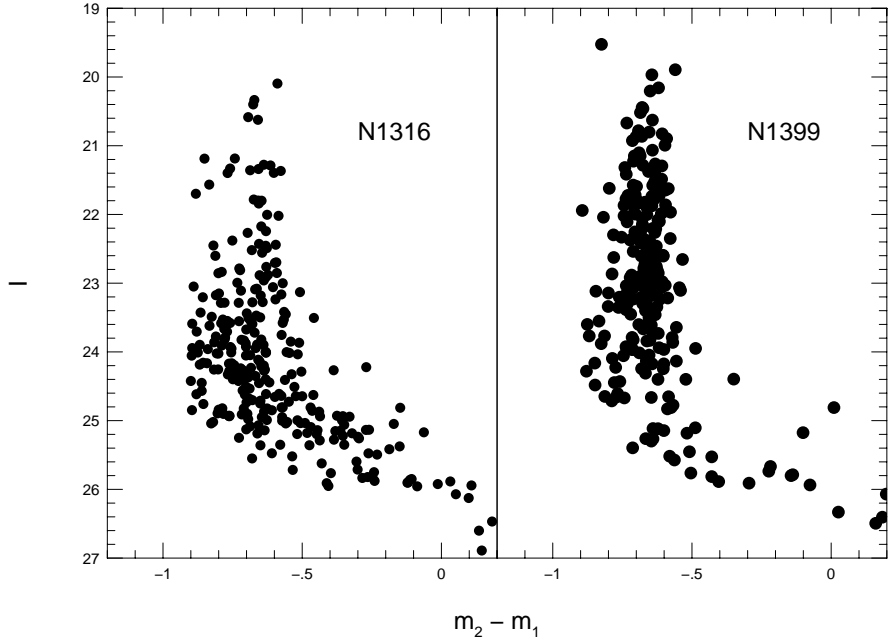}
\caption{The distribution of $m_2 - m_1$, the difference in
magnitude between apertures of radius one pixel and apertures of
radius two pixels, as measured in the F814W frames.  The clusters in
NGC 1399 have been sampled from the same portions of the CCD frames
used for the unobscured clusters in NGC 1316. Note that a cutoff of
$m_2 - m_1 = -0.9$ (-1.1 for the PC frames) has already been
imposed to excise as many background galaxies from the sample as
possible.  \label{fig:m1m2}}
\end{figure}

In Figure \ref{fig:m1m2} we show the distribution of clusters over
$m_1 - m_2$ (the difference between magnitudes measured for one- and
two-pixel-radius apertures) and magnitude for NGC 1316's unobscured
clusters and for NGC 1399 globular clusters which fall in the same,
outer portions of the WFPC2 detectors. (By limiting our comparison to
the same regions of the detectors we minimize the effects that
position-dependent changes in the shape of the PSF might have on
object centering.) The mean measured $m_1 - m_2$ for objects with $I <
25$ are identical, but the overall distribution of $m_1 - m_2$ is
about 40\% broader for NGC 1316 than it is for NGC 1399.  Whereas a
Kolmogorov-Smirnov test shows that the distributions of $m_1 - m_2$ of
clusters in NGC 1399 and NGC 1404 cannot be distinguished at the 50\%
confidence level, comparing NGC 1399 and NGC 1316 using the same test
indicates that we can reject the null hypothesis at $>99.8\%$
confidence level.

The distribution of colors, magnitudes, and $m_1 - m_2$ in NGC 1316 is
in some respects similar to that of compact field objects in F0336,
but the total surface density of objects is some 15 times larger.
Disregarding the possibility that there may be an undiscovered,
populous cluster of galaxies situated beyond NGC 1316 which mimics the
color and luminosity distribution of field galaxies, it is exceedingly
unlikely that the observed number of objects in NGC 1316 could be a
chance fluctuation in the density of background galaxies. Moreover,
over the limited, dust-free region available in the WFPC2 field of
view, the radial distribution of the unobscured clusters with $B <
26.5$ follows a powerlaw form with $\alpha \approx -1.5 \pm 0.5$.
This is not significantly different from NGC 1316's surface brightness
profile, which Shaya et al. (1996) found to have an asymptotic slope of
$-1.16 \pm 0.02$. This is consistent with the radial distributions of
many other cluster systems and supports our view that these objects
must be physically associated with NGC 1316.

Given the recent merger which has clearly taken place in NGC 1316, one
might have expected significant numbers of metal-rich {\it globular}
clusters to have formed as prescribed by the model of Ashman \& Zepf
(1992). However, neither we nor Shaya et al. (1996) (who also studied
objects in the dust-obscured central regions) have found evidence for
presumed young globular clusters of the type found by Holtzman et al. 
(1992) in NGC 1275 and Whitmore et al. (1993) in NGC 7252.  This may be
due to the much smaller amount of gas and dust in NGC 1316 compared to
these other galaxies.  Alternatively, perhaps the active nucleus and
twin jets of NGC 1316 are preventing the condensation of gas clouds
sufficiently large to produce globular clusters. Inspection of the
Northeastern quadrant of NGC 1316 in Figure \ref{fig:plate2} reveals
structures in the dust which are reminiscent of the photoevaporating
columns of M 16 (Hester et al. 1996), though on a much larger scale. The
radially-oriented columns in the dust could be a result of erosion by
a radiation or particle flux originating in the central regions of the
galaxy. Alternatively, we may be seeing Rayleigh-Taylor instabilities
in action.  NICMOS observations of the tips of these columns may be
useful in determining whether proto-open clusters are indeed
condensing out of the material.

Given that the only large open cluster populations we know of are in
spiral galaxies, would the existence of large numbers of open clusters
in NGC 1316 imply that the merger remnant we see today must at one
time have been a sizable spiral? This would certainly be consistent
with the presence of dust, though the relatively small amount of dust
compared with the amount seen in NGC 1275 suggests that NGC 1316's
intruder may have been a smaller or an earlier type of spiral. Could
it be that many early-type galaxies have exponential cluster
luminosity functions (or at least significant exponential components),
but that they have only really become detectable with the the advent
of the {\it HST}? If the objects in NGC 1316 really are old open
clusters, where are the globular clusters?  If NGC 1316 had itself
also been a spiral galaxy before the merger, then a globular cluster
specific frequency of $\sim 0.5$ might not seem so unreasonable.
However, a merger between two spiral galaxies generating virtually no
new globular clusters would be at odds with the model predictions of
Ashman \& Zepf (1992).  Could it be that the clusters we see in NGC
1316 started out as all cluster systems do, but that they have somehow
avoided the tidal shocking and disruption which may be responsible for
the log-normal GCLF seen in most other galaxies (Vesperini 1998)?
Given the anomaly that NGC 1316's cluster system represents, these
questions are well worth pursuing and may provide us with an important
key to understanding both star cluster and galaxy formation.

\section{Intergalactic Globular Clusters. \label{sec:inter}}

Theuns \& Warren (1996), Arnaboldi et al. (1996), and Ferguson et al. 
(1997) recently detected significant numbers of stars and planetary
nebulae at rather large distances from the cD galaxies NGC 1399 and M
87. Grillmair et al. (1994a) found that globular clusters around NGC
1399 seem to be kinematically related more to the whole Fornax cluster
than to the central galaxy itself. This has prompted us to examine our
data for evidence of such intergalactic nomads among the globular
clusters in our sample.  The various components of the Fornax cluster
(stars, globular clusters, and cluster galaxies) all seem to fall off
with distance from NGC 1399 as $f \propto r^{-1.5}$ (\eg Grillmair
et al. 1994a; Kissler-Patig et al. 1997a). In the WF frames only, we
find 517, 74, and 48 objects with $B < 26.5$ in fields N1399, F0338,
and F0336, respectively.

The radial surface density distributions of globular clusters in cD
galaxies generally have a fairly extended ``core" (Grillmair et al.
1986; Lauer \& Kormendy 1986; Grillmair et al. 1994b) where the
surface density becomes nearly constant.  If we thus ignore our
central pointing and assume that in the outer parts the surface
density falls off as $r^{\alpha}$, then solving

\begin{equation}
f = a r^{\alpha} + b 
\end{equation}

\noindent for $\alpha = -1.5$ and using only the outer fields at 0.14
and 1.4 degrees, we find $b = 10.4 \pm 1.5$ arcmin$^{-2}$. Thus the
surface density of unresolved objects (10.6 arcmin$^{-2}$) at 1\fdg4
from NGC 1399 would be entirely consistent with a roughly constant
surface density of compact background galaxies. Put another way, under
the assumption that the globular cluster surface density profile
maintains a power-law form with $\alpha = -1.5$ at all radii, we would
predict only $0.8 \pm 0.3$ globular clusters in our outermost field.
[As an aside, this powerlaw would predict $750 \pm 170$ globular
clusters in our NGC 1399 field, or about 60\% more than we actually
see, which is consistent with the degree of core-flattening in the
surface density distribution found by Kissler-Patig et al. (1997a)].
If we adopt the flatter slope with $\alpha = -1.2$ found by Forbes et
al. (1998), we would predict $1.7 \pm 0.7$ globular clusters in our
outermost field.  Clearly, the data are consistent with no
intergalactic globular clusters at 1\fdg4, but are insufficient to
rule them out.

Despite the relatively large color/metallicity spread in NGC 1399's
globular cluster system, Figure \ref{fig:cm} demonstrates how well
defined their distribution is over color and magnitude compared to the
background field. Indeed, one could draw a box with $22 < B < 27$ and
$1.5 < B - I < 2.3$ and capture more than 95\% of the cluster sample.
If we then place such a box over the color-magnitude distribution of
objects in F0336, we find a total of 7 objects crowding the very faint
end of the box.  Close visual inspection of these objects reveals that
two of the seven objects are probably galaxies, but we cannot rule out
that the remaining five objects may be globular clusters in the
envelope of NGC 1399. However, if this is so, then the color-magnitude
distribution of these outlying clusters must be very different from
those near the center of the galaxy; the odds of obtaining the
observed color-magnitude distribution by sampling the GCLF in Figure
\ref{fig:lf} are vanishingly small.

\section{Summary and Conclusions. \label{sec:summary}}

We have analyzed WFPC2 images of three early-type galaxies and two
background fields in the Fornax cluster. From an investigation of the
global distribution over color and magnitude, we conclude the
following:

\begin{itemize}

\item{} The color distributions of globular clusters in the central
regions of NGC 1399 and its nearby neighbor NGC 1404 are bimodal,
and statistically very similar to one another.

\item{} With respect to the color distributions seen in NGC 1399 and
NGC 1404, objects in a field 8.5\arcmin~ from NGC 1399 have a
significant blueward bias. The metallicity gradient found by Ostrov et
al. (1993) in NGC 1399 appears to be present even among the faintest
clusters.

\item{} The luminosity functions of globular clusters in NGC 1399 and
NGC 1404 are essentially identical to one another and very similar to
the luminosity function of globulars in M 87. 

\item{} The luminosity function for the blue globulars in NGC 1399
peaks $0.36 \pm 0.12$ magnitudes brighter in $B$ than the luminosity
function of the red clusters.

\item{} The luminosity function of objects surrounding NGC 1316
is unlike that found for globular clusters in the other galaxies. Based on
the bluer color distribution, the exponential shape of the luminosity
function, and the relatively large spread in cluster sizes, we tentatively
identify these objects with an extensive population of old {\it open}
clusters. 

\item{} The measured peak of NGC 1399's luminosity function (converted
to $V$-band magnitudes) is 0.01 $\pm$ 0.1 magnitudes fainter than that
of M 87, yielding $H_0$ = 82 $\pm$ 8 km s$^{-1}$ Mpc$^{-1}$, where the
uncertainty reflects only the effects of random errors.

\item{} We see no evidence for intergalactic globular clusters, though
our field of view is too small to strongly constrain this finding.

\end{itemize}

\acknowledgments

CJG is grateful to S. Faber and J. Kormendy for many interesting and
useful discussions. We also thank M. Kissler-Patig for a critical
reading of the manuscript. Finally, we are grateful to an anonymous
referee for a very helpful report. This research was supported in part
by STScI Grant No. GO-05990.01-94A, and by faculty research funds from
the University of California, Santa Cruz.

\end{document}